# Redispersible Hybrid Nanopowders:
## Cerium Oxide Nanoparticle Complexes with Phosphonated-PEG Oligomers

L. Qi[1], A. Sehgal[1], J.-C. Castaing[1], J.-P. Chapel[1,@], J. Fresnais[2], J.-F. Berret[2], F. Cousin[3]

*(1) Complex Fluid Laboratory, UMR CNRS/Rhodia 166, CRTB Rhodia Inc., 350 G. Patterson Blvd, Bristol PA, 19007*
*(2) Matière et Systèmes Complexes, UMR 7057 CNRS Université Denis Diderot Paris-VII, Bâtiment Condorcet, 10 rue Alice Domon et Léonie Duquet, 75205 Paris, France*
*(3) Laboratoire Léon Brillouin, UMR CEA-CNRS 12, CEA- Saclay, 91191 Gif-sur-Yvette, France.*

**Abstract :** Rare earth cerium oxide (ceria) nanoparticles are stabilized using end-functional phosphonated-PEG oligomers. The complexation process and structure of the resulting hybrid core-shell singlet nanocolloids are described, characterized and modeled using light and neutron scattering data. The adsorption mechanism is non-stoichiometric, yielding the number of adsorbed chains per particle $N_{ads}$ = 270 at saturation. Adsorption isotherms show a high affinity of the phosphonate head for the ceria surface (adsorption energy $\Delta G^{ads}$ ~ -16 kT) suggesting an electrostatic driving force for the complexation. The ease, efficiency and integrity of the complexation is highlighted by the formation of nanometric sized cerium oxide particles covered with a well anchored PEG layer, maintaining the characteristics of the original sol. This solvating brush-like layer is sufficient to solubilize the particles and greatly expand the stability range of the original sol (< pH 3) up to pH = 9. We underscore two key attributes of the tailored sol: i) strong UV absorption capability after functionalization and ii) ability to re-disperse after freeze-drying as powder in aqueous or organic solvents in varying concentrations as singlet nanocolloids. This robust platform enables translation of intrinsic properties of mineral oxide nanoparticles to critical end use.

## I – Introduction

The size–dependent properties of nanoparticles have generated diverse scientific interests. Their high surface area, shape, surface chemistry and intrinsic properties (dielectric constant, radiation absorption, sensitivity to external field…) are currently fueling innovations and driving technological breakthroughs in many areas. These range from biotechnology[1,2], optics[3], photonics[4], electronics[5], energy[6,7] to materials science[8]. Numerous synthesis routes of inorganic nanoparticles now exist in the literature and availability of small (< 10 nm), fairly monodisperse and non-aggregated inorganic nanoparticle of various chemistries is no longer a constraint. The primary limitation of "solubility" or inorganic nanosol stability to perturbation from their "as synthesized" state during processing represents the critical challenge for any given application. The sols are extremely sensitive to changes in their physico-chemical environment such as pH, ionic strength, temperature and concentration often leading to aggregation. The need to offset the attractive *van der Waals* interaction that drives the aggregation of the particles is particularly significant for dense nanoscale inorganic particles with low interparticle minimum radial separations. This has been addressed through the adsorption of an organic layer (corona, shell or adlayer) around the particle promoting an electro-steric or steric stabilization of the sol. This complexation is prevalent in organic solvents in the absence of charge. In addition to stabilization, the organic shell and/or the presence of some specific groups in the corona may also confer some functionality to the high surface area of the nanoparticles. Surface derivatization is crucial in biomedical applications like drug delivery, immunoassay or cell imaging where the control of the interaction between the nanoparticles and bio-macromolecules, cells or living tissues drives toxicity[1,9,10].

Metal-oxide nanocrystals such as cerium ($CeO_2$), iron ($Fe_2O_3$), zirconium ($ZrO_2$) and titanium ($TiO_2$) oxides represent a particularly important class of inorganic nanoparticle systems widely used in the diverse fields above. Despite this diversity in chemistry and application these particles share some common features. Their synthesis usually occurs in very acidic (or basic) environments by precipitation of metallic salts. The result is a stabilized sol at low pH (typically < pH 2) via electrostatic repulsion between particles. The main drawback is a very acute sensitivity to any variation of physicochemical conditions in the aqueous solution. In order to increase their stability range, low molecular weight molecules or ligands like citric acid[11] or poly(acrylic acid)[12] are generally used. The COOH groups of the ligands form complexes with the surface hydroxyl groups of the nanoparticles. At sufficiently high pH where the carboxylic groups are ionized (> pH = 5), electrostatic stabilization of the sol occurs. The fundamental physical and chemical characteristics underlining the formation of the complexations are not straightforward since the complexations require some mechanical stress (shear, high energy ultrasounds)[13,14] to re-disperse the sol or a two step process (low pH aggregation - high pH re-dispersion[12]). The electrosteric repulsion conveyed via the ligands (eg. carboxylic groups) though less sensitive to perturbations of pH and/ or ionic strength compared to the bare sols is still ionic and serves only to extend utility for limited application. Alternatively, the use of head-tail architecture has been shown recently to be an efficient strategy for the dispersion of alumina nanoparticles[15] (PEG tail and a gallol anchoring group) or the surface modification of yttrium oxide nanoparticles[16].

We report a facile complexation process giving rise to robust and versatile hybrid metal oxide nanoparticle sols with a well anchored neutral corona for wide applicability. This process obviates the necessity for specialized processing steps as precipitation–redispersion. The choice of the complexing group(s), anchored on the particle surface, and the nature of the oligomeric tail, in the solvent, is of paramount importance. Among various anionic moieties it has been shown that phosphonic acid groups bind strongly to a variety of metal oxide surfaces at room temperature in aqueous solution[16,17].

We also chose polyethylene glycol (PEG) groups that are widely used in biomedical applications to prevent non-specific adsorption of proteins[18]. An end grafted PEG layer may provide some biological stealth to the hybrid metal oxide nanosols[19].

Along these lines, we highlight the potential and versatility of phosphonic acid terminated PEG oligomers to functionalize rare earth oxide nanoparticles. Cerium oxide is chosen to illustrate this approach due to its growing importance in science and technology ranging from material science (catalysis, polishing, optics…)[20-24] to biomedical[25-27] applications.

In the following section, we describe, characterize and model the complexation process and the resulting nanostructure of the hybrid particles. We then underscore two key attributes of the tailored sol: i) strong *UV absorption* capability after functionalization and ii) ability to *re-disperse* after freeze-drying as powder in *aqueous or organic solvents* in varying concentrations as singlet nanocolloids.

## Results and discussion

### *Light scattering*

To understand and model the adsorption of the oligomers onto the nanoceria particles, hybrid solutions are prepared as a function of the volume ratio X defined in equation 7 with an overall concentration $c$ kept constant. The Rayleigh Ratios and hydrodynamic diameters are measured and plotted versus the mixing ratio X (*Figure 1*). $X = 10^{-4}$ corresponds to a solution containing PPEG oligomers only. $X = 1000$ corresponds to a solution containing bare nanoceria particles only. When X decreases from 1000 to 0.01, the size of the coated nanoceria increases gradually up to a critical value noted $X_P$ where it then saturates around $D_H = 13$ nm. We interpret this result as the progressive coating of the particles by the oligomers until full coverage occurring at $X_P$. By further decreasing X, the number of fully covered particles decreases until their number is too low to be detected by DLS (around X ~ 0.01). The normalized Rayleigh Ratio (normalization by the Rayleigh Ratio of the bare nanoceria $R_{NP}(X = \infty)$, as measured by SLS, is also monitored as a function of X. Here as well, one can clearly identify a critical ratio X equal to $X_P$ where the Rayleigh Ratio starts to decrease progressively with X. We interpret this result as the dilution of the fully covered particles by the free oligomers present in the bulk, in agreement with DLS results.

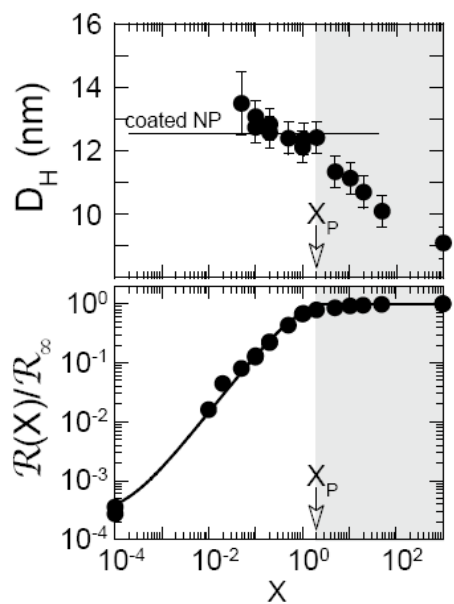

**Figure 1.** *Raleigh ratio and hydrodynamic radius $D_H$ vs. X at low pH (pH = 1.5). The continuous line is a fit to the data using a non-stoichiometric model for adsorption (see the analysis section). In the white region the fully covered particles coexist with free PPEG oligomers.*

To confirm the existence of a critical volume ratio at $X_P$ below which particles are fully covered, the pH of the solutions is raised (it should be recalled that bare nanoceria solutions start to aggregate at pH = 3 and precipitate above pH = 5)[28]. At pH = 9, a solution made at $X < X_P$ remains stable suggesting that the steric repulsion generated by the PEG layer is sufficient enough to prevent the destabilization of the sol. This result is confirmed by a Rayleigh Ratio equal to the value found at pH = 1.5. Above $X_P$, as expected, the solution begins to aggregate due to the lack of oligomers on the particles resulting in a very low steric hindrance.

Furthermore, a comprehensive description of the behavior of the coated particles as a function of *pH* and *mixing ratio X* is undertaken. A large number of different experimental conditions are gathered in Figure 3 representing the whole stability phase diagram of the $CeO_2$-PPEG hybrid nanoparticles. As the pH is progressively increased, the solutions become destabilized at pH > 10.5. This critical upper stability limit is identical to the $pK_a$ of the (110) and the (111) facet of the ceria nanocrystal unit cell when the charge changes from zwitterionic to negative[28]. We postulate desorption of the phosphonated-PEG at this pH due to the co-incidence of the transition to the aggregated state (see section 3.3). Another feature of this phase diagram is the existence of a general critical $X = X_P$ For all values of $X > X_P$, the particle size increases progressively at a lower pH leading to early phase separation.

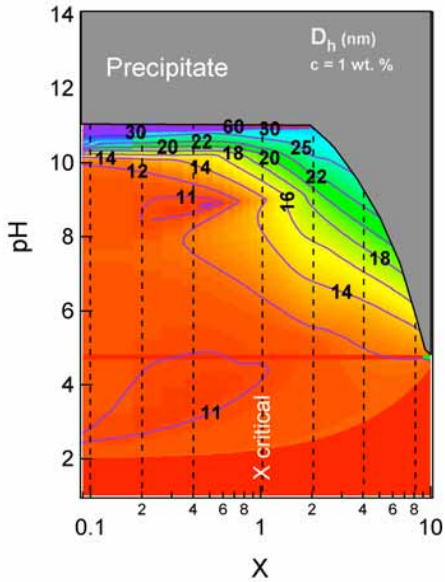

**Figure 2.** Stability phase diagram of the PPEG-$CeO_2$ nanoparticles: hydrodynamic radius $D_H$ vs. X and pH. The grey region represents visible phase separation. Contours and colors (red to blue) indicate $D_H$ variation from 10 nm to 40 nm.

To obtain more quantitative results on this system, the Rayleigh Ratios were measured at pH = 1.5 for both $CeO_2$ and $CeO_2$-PPEG dispersions as a function of the $CeO_2$ weight fraction (
Figure *3*).

The two slopes are different because the Rayleigh Ratio is proportional to K and $M_w$, as shown in equation 11. The slopes of the linear regressions for the $CeO_2$-PPEG and $CeO_2$ Rayleigh ratios versus the $CeO_2$ weight fraction are 0.1284 and 0.0743, respectively. The clear difference in the 2 slopes gives us the possibility to compute precisely the number of adsorbed PPEG chains per nanoceria. Using expression (1) described in a former publication[12] and values from Table 1 one finds 270 ± 30 phosphonate-PPEG per $CeO_2$ particle.

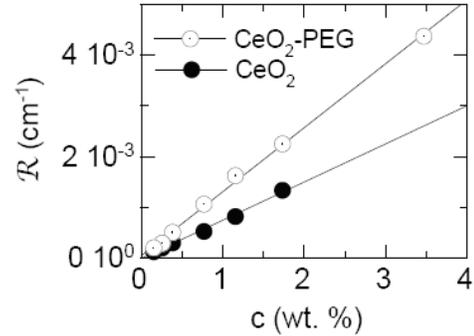

**Figure 3.** Rayleigh Ratios of $CeO_2$ and $CeO_2$-PPEG solutions versus the $CeO_2$ weight fraction.

$$\frac{R^{CeO_2-PPEG}}{R^{CeO_2}} = \frac{K^{CeO_2-PPEG}}{K^{CeO_2}}\left(1 + n_{ads}\frac{M_W^{PPEG}}{M_W^{CeO_2}}\right)^2$$
(1)

The molecular weight of a hybrid nanoparticle is hence $M_w$ = 440 ± 30 Kg/mol. The organic layer represents 38 % of the total weight.

| Specimen | dn/dc (cm³.g⁻¹) | K (632.8nm) (cm².g⁻²) | $M_w$ (g.mol⁻¹) | $D_H$ (nm) | $R_g$ (nm) |
|---|---|---|---|---|---|
| PPEG | 0.102 | 7.58 10⁻⁸ | 624 | — | — |
| $CeO_2$ | 0.192 | 2.69 10⁻⁷ | 269000 | 9.2 | 3.2±0.1[a] |
| $CeO_2$-PPEG | 0.154 | 1.72 10⁻⁷ | 439000 | 12.3 | 4.4±0.1[a] |

**Table 1** : Molecular weight ($M_w$), refractive index increment (dn/dc), hydrodynamic diameter ($D_H$), light scattering coupling constant K (at 6328 nm), characterizing PPEG oligomer, the nanoceria, and the PPEG coated nanoceria investigated in the present work. The gyration radii ($R_g$) were measured by SANS[a].

The determination of the mutual diffusion coefficients for both dispersions at different weight fractions allows for the calculation of $D_0$, the self-diffusion coefficient of the bare and coated nanoparticles (Figure 4). Careful DLS measurements are able to distinguish a different behavior for coated and non coated particles. Using the Stokes-Einstein relation, the hydrodynamic diameters $D_H$ of bare and PPEG coated nanoparticles were found to be $D_H$ = 9.2 nm and $D_H$ = 12.3 nm respectively. One can estimate a thickness of 1.6 nm for the PPEG corona. The theoretical fully extended contour length of the PPEG molecules is about 4.5 nm. The difference expresses the grafting density of the PPEG onto the particle (270 chains per particles or 1 chain/nm$^2$) typical for oligomers adsorbing onto nanoparticles[15] (the non spherical shape of the particle might also play a role). A monolayer of PPEG densely grafted (and fully stretched) would give a grafting density of 4 chains/nm$^2$ PPEG. The non-fully stretched configuration of the PPEG is however sufficient to provide a very good steric stabilization.

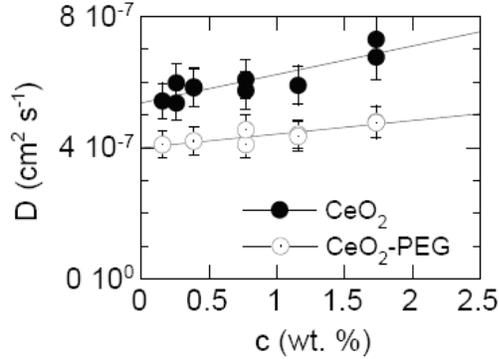

**Figure 4.** *Mutual diffusion coefficient versus $CeO_2$ weight fraction in $CeO_2$ and $CeO_2$-PPEG dispersions. The extrapolation of the straight lines at c = 0 gives $D_0$.*

## Non-stoichiometric model

In order to model the adsorption of the PPEG oligomers onto the particles two different mechanisms that generally describe how polymers coat nanoparticles have been considered. The mechanisms named stoichiometric (ST)[29] and non stoichiometric (NST) models[30] (Figure 5) are described in the literature. The main difference between them only appears above $X_P$ ($X > X_P$) where the particles are no longer fully covered (shortage of oligomers). Two distinct mechanisms are likely. First, the coated nanoparticles coexist with non coated ones giving a number of oligomers/particle $n_{ads}$ constant throughout the X range (ST model). Second, all nanoparticles are equally covered (NST model) and $n_{ads}$ is hence a function of X. Below $X_P$ ($X < X_P$) where the two models merge, the coated nanoparticles coexist with free oligomers present in the bulk solution.

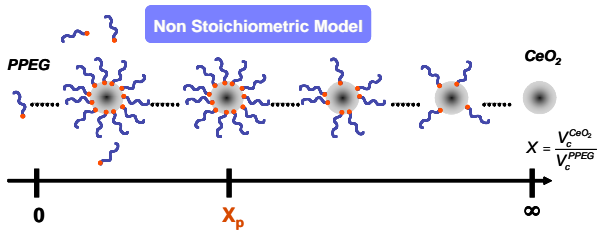

**Figure 5.** *Schematic representation for the non-stoichiometric adsorption mechanism of phosphonated-PEG onto nanoceria particles. Below the critical volume ratio $X_P$, the particles are fully covered and in equilibrium with a decreasing number of non-absorbed oligomers. Above $X_P$, the number of grafted oligomers progressively decreases toward bare particles (X = ∞). At $X_p$, all the oligomers present in the initial solution are adsorbed onto the particle.*

From a closer look to the data of Figure 1, and within the experimental error of the light scattering technique, $D_H$ appears quite constant (around 13 nm) up to a value of X close to $X_P$ when it starts do drop smoothly until it reaches the value of the bare nanoceria particle (9.2 nm) at X = ∞ (extreme point on the right). This behavior suggests a NST model for the adsorption. In addition, no precipitation above $X_P$ is seen in the phase diagram (Figure 2) for pH < 4. This result is in contradiction with a ST model that would predict the coexistence between bare and coated particles leading to the aggregation of a portion of the sol. Therefore, a non-stoichiometric adsorption mechanism is selected to model the oligomer adsorption onto the nanoparticles. In what follows, the X-dependence of the scattered intensity will be modeled accordingly.

In the experiments described previously, the total concentration $c$ of active matter ($CeO_2$ + PPEG) is kept constant. The respective nanoparticles and oligomers concentration are given by equation 7 regardless of any interaction between them. One can define the nominal number of oligomers/particle as $n_{ads}(X) = M_w^{CeO_2} / X M_w^{PPEG}$. Below $X_P$, $n_{ads}$ is equal to $n_{ads}(X_p)$ and the solution contains fully coated nanoparticles ($CeO_2$) and free oligomers (PPEG). Above $X_P$, $n_{ads}$ is a function of X and the solution contains only partially covered particles. The total scattering intensity expresses hence as the sum of different contributions. As a result, the Raleigh ratio normalized to its value at X = ∞ ($R = K^{CeO_2} M_W^{CeO_2} c$) reads as

$$\tilde{R}(X < X_p) = \tilde{K}M\left(\frac{X_p - X}{X_p(1+X)}\right) + \left(\frac{\tilde{K}^{1/2} + X_p}{X_p}\right)^2 \left(\frac{x}{1+x}\right) \quad (2)$$

The first term arises from the free oligomers present in the bulk and the second term to the fully covered particles. Similarly, at high X the contribution arises solely from the partially covered particles and the intensity goes as

$$\tilde{R}(X > X_p) = \left(\frac{1+X}{X}\right)\left(\frac{\tilde{K}^{1/2} + X}{1+X}\right)^2 \quad (3)$$

where $\tilde{K} = \dfrac{K^{PPEG}}{K^{CeO_2}}$ and $\tilde{M} = \dfrac{M_W^{PPEG}}{M_W^{CeO_2}}$.

When X goes either to zero (only oligomers in the solution) or to infinity (only nanoparticles), the ratio tends, as expected, to $\tilde{K}\tilde{M}$ or 1, respectively. The normalized Raleigh ratios do not depend on the total concentration $c$; therefore the expression should be valid at all concentration in the dilute regime. Equations 2 and 3 are used to fit the scattering data of *Figure 1* with $X_P$ as the sole adjustable parameter. All other quantities such as the coupling constant $\tilde{K}$, the molecular weights of both oligomers and nanoparticles are known. The result of the fitting is shown in *Figure 1* as a solid curve. The agreement between the model and the data is excellent. $X_P$ is found to be 1.65 leading to a number of oligomers/particle $n_{ads}(X_p) = 260 \pm 30$ in close agreement with the results found through the data of
Figure 3.

## SANS scattering

To complement light scatterings data, SANS is performed on both bare and PPEG-coated nanoparticles to investigate the nanostructure of the core-shell hybrid system (Figure 6). The scattered intensity I(q) of spherical objects can be decomposed into a structure factor S(q) and form factor P(q). At sufficiently low concentration (< 1wt. %), the structure factor is equal to 1 (no interparticle interaction), the scattered intensity is proportional to the concentration and the q-dependence of the intensity reflects the form factor P(q) of the aggregates. Figure 6 shows the form factor of both bare and modified particles. Because the measurements of the scattered intensity are performed in $H_2O$ and $D_2O$ for the $CeO_2$ and PPEG-$CeO_2$ system respectively, the curves at high q can not be superimposed. The extra contribution to the scattered intensity due to the presence of the organic layer is, however, clearly seen at low q in the Guinier representation where $R_g q < 1$ (inset Figure 6). Here, the logarithm of the intensity decreases linearly with $q^2$ and from the straight lines, we deduce a radius of gyration $R_g = 3.2 \pm 0.1$ nm for the bare nanoparticles and $R_g = 4.4 \pm 0.1$ nm for the PPEG coated ones. For homogeneous and monodisperse spheres of radius R, $R = R_H = 1.29\, R_g$. In the present case, for the bare particles, we find $R_H/R_g \sim 1.44$. The discrepancy likely reflects the polydispersity in size of the cerium oxide particles. In the Porod representation ($I(q) \times q^4$, not shown here), the first oscillation of the form factor shifts toward lower q after the coating (from 0.11 to 0.07) supporting the presence of the organic layer. However, due to polydispersity it is difficult to extract the radius R of the particles ($q^{peak} R = 4.57$). In addition, from experiments performed with solutions having different $H_2O/D_2O$ volume ratios it is possible to measure the average scattering length density ($\tilde{\rho}$) of the bare and coated particles. The plot of $\sqrt{I(q \to 0)/c}$ as a function of the proportion of $H_2O$ in the ($H_2O+D_2O$) mixed solution gives a straight line (not shown here). The ratios that cancel out the intensity correspond to $\tilde{\rho}$. We find $\tilde{\rho}_{CeO_2} = 4.99\ 10^{10}$ cm$^{-2}$ for the bare particles. In the case of coated ones the ratio gets rather through a minimum (because it is not possible to match jointly the particle and the organic shell) giving $\tilde{\rho}_{CeO_2-PPEG} = 1.93\ 10^{10}$ cm$^{-2}$.

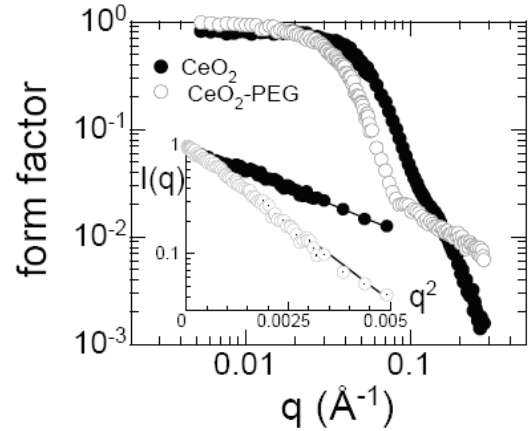

*Figure 6.* SANS form factor P(q) for bare and coated nanoparticles in double logarithmic scale. Inset: Guinier representation of the intensity for the same samples (I(q) versus $q^2$). From the straight lines, the gyration radii for bare and coated nanoceria were calculated, $R_g$ = 3.2 nm and 4.4 nm, respectively.

From the results gathered so far, the coated nanoparticles seem to have a hybrid core-shell structure. In that case, the intensity excess as $q \to 0$ due to the PPEG shell can be approximated by[12]

$$\left(\dfrac{I^{CeO_2-PPEG}}{I^{CeO_2}}\right)_{q\to 0} = \left(1 + n_{ads}\dfrac{\rho^{PPEG} - \rho^{D_2O}}{\rho^{CeO_2} - \rho^{D_2O}}\dfrac{v_M^{PPEG}}{v_M^{CeO_2}}\right)^2 \quad (4)$$

The ratio on the left side of equation 4 was found experimentally. The intensity scattered by the coated ceria was 38 times that of bare particles in $D_2O$. Knowing $\tilde{\rho}_{CeO_2}$, the molar volume of bare ceria $v_M^{CeO_2}$ can also be calculated from the intensity scattered as $q \to 0$ of a diluted ceria solution. The molar volume of a PPEG oligomer is estimated via its $M_w$ and a bulk density of 1.13 ($v_M^{PEG} \sim 910$ Å$^3$). The values of $\rho^{D_2O}$ (= $6.33\ 10^{10}$ cm$^{-2}$) and $\rho^{PEG}$ (= $0.84\ 10^{10}$ cm$^{-2}$) are classically computed according to their composition and density. Finally, one ends up with a number of adsorbed PPEG oligomers/particle $n_{ads}$ = 230 ± 30.

Both light and neutron scattering data are in very good agreement, confirming the consistency of our approach and the accuracy of the results. From quantitative scattering data we are able to deduce the main features of the core-shell nanostructure of the hybrid particles. The integrity and efficiency of the coating process is also highlighted. The PPEG functionalization does not alter the main specificities of the original sol e.g. singlet nanometer size cerium oxide particles and greatly expands its stability range up to pH = 9.

**Adsorption isotherms**

In order to evaluate the affinity of the phosphonated-PEG oligomers with the surface of cerium oxide nanoparticles, adsorption isotherms are measured on macroscopically flat $CeO_2$ model surfaces with the help of optical reflectometry. As seen in Figure 7, in both values of pH investigated the curves present a rather sharp increase of the adsorbed amount at very low concentration. This rapid flattening out of the adsorbed amount suggests a rather high affinity of the oligomers toward cerium oxide surface. The data of Figure 7 can be fitted using a Langmuir [15] model where the oligomers adsorbed amount $\Gamma(c)$, specifically

$$\Gamma(c) = \frac{k\Gamma_{sat}c}{1+kc} \qquad (5)$$

where $\Gamma_{sat}$ is the maximum adsorbed amount, $c$ is the (equilibrium) concentration of oligomers in the aqueous solution and $k$ is the adsorption constant ( = $\frac{k_{adsorption}}{k_{desorption}}$). The fit gives values of $\Gamma_{sat}$ = 1.8 mg/m², $k$ = 1.6 $10^5$ l/mol and $\Gamma_{sat}$ = 0.8 mg/m², $k$ = 1.4 $10^5$ l/mol for pH = 1.5 and pH = 6.5 respectively. In both case $k$ is quite large leading to small values for $k_{desorption}$, consistent with a high affinity type of isotherm. The magnitudes of the adsorbed amounts are typical for organic matter adsorbing onto inorganic surfaces. The adsorbed amount (at pH = 1.5) computed from light scattering results ($D_H$ = 9.2 nm, 270 oligomers/NP) is equal to 1.1 mg/m² (assuming a spherical shape for the NP). Hence, both approaches agree reasonably well. In addition, due to a size comparable to a nanoparticle, an oligomer adsorbing on top of the ceria model surface does not see a homogeneous layer, but rather a compact set of nanoceria spheres with at most a 90 % coverage (in the case of an hexagonal close-packing). This layer develops a surface area of $2\pi R^2$ × number of particles (the surface area of hemispheres of radius R). The renormalized adsorbed amount is ~ 1 mg/m² in very good agreement with the value found through light scattering experiments.

The free energy of adsorption $\Delta G^{ads}$ of the PPEG oligomers on the ceria surface is estimated from the following expression:

$$\Delta G^{ads} = -k_B T \ln(\frac{k}{V_{water}^m}) \qquad (6)$$

where $k_B$ is the Boltzmann constant, $T$ is the temperature, and $V_{water}^m$ is the molar volume of water (0.018 L/mol)[31,32]. The adsorption free energies are found to be -16 $k_B T$ and -15.8 $k_B T$ for pH = 1.5 and pH = 6.5 respectively. Though the exact nature of the bond between PPEG and ceria is not known at this stage, the measured free energies might indicate an adsorption mechanism controlled by electrostatic interaction[15]. At pH = 1.5 the ceria nanoparticles are cationically charged ($CeO-H_2^+$) and the PPEG are only slightly anionic (3%, $pK_1$ ~ 2.7). However, it has been reported [33, 34] in the literature that the presence of oppositely charged nanoparticles can affect considerably the acid-basic properties of weak poly-acid by making group or chain ionization easier. Although we do not know at this stage how to quantify such mechanism, it can possibly apply in our system and explain the presence of a larger charge on the weak phosphonic acid at low pH leading to the strong adsorption measured.

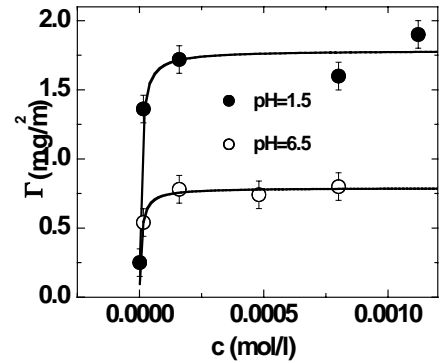

*Figure 7. Adsorption isotherms $\Gamma(c)$ of PPEG oligomers onto model $CeO_2$-coated substrates at low (1.5) and neutral pH (6). The solid curve is the Langmuir fit to the data.*

At pH = 6.5, close to its IEP, ceria nanoparticles are less cationic and PPEG oligomers are 50% anionically charged ($pK_2$ ~ 7.8). The lower amount measured at higher pH may be due to electrostatic repulsion between the charged oligomers while adsorbing onto the surface. At this pH, the presence of more than one charged group in the phosphonated head could possibly force the oligomers into a flatter configuration leading to a lower adsorbed amount as well. The acid-basic interaction between the hydroxyl groups (Ce-OH) of the ceria and the remaining phosphonic acid groups (O = P-(OH)) of the PPEG might also favor adsorption leading possibly to a Ce-O-P = O bond with the release of a $H_2O$ molecule. The reactivity of phosphonic acid groups toward metal oxide surfaces is known to correlate with its basicity[35]. In the case of yttrium oxide ($Y_2O_3$) nanoparticles (yttrium and cerium are rare earth elements with similar electronegativity and ionic radii)[36] the reaction (''esterification'') takes place at room temperature[16]. Furthermore, the modified $CeO_2$ sol was able to withstand a week-long dialysis against pure water without aggregating, highlighting the robustness of the organic coating. It should be noted finally that at very high pH (> 11), the sol precipitates, as can be seen in the phase diagram of Figure 2 indicating possible hydrolysis of Ce-O-P bonds in the aggressive alkaline environment as the ceria surface changes character from zwitterionic to anionic[28]. The adsorption results give evidence of the specificity of a phosphonated oligomer to create a robust metal oxide hybrid sol.

### Benefits: UV absorption and Redispersible Nanopowders

The data and models offer a clear insight into the formulation process and the nanostructure of the modified hybrid nanoparticles, and translate into two valuable bulk properties of the new sol: UV absorption and the creation of redispersible nanopowders by freeze drying the nanosol.

One of the well known characteristic properties of cerium oxide particles is strong UV absorbance with minimal absorbance in the visible regime. However, given the large surface area once covered with an organic layer and the possibility to form metal-ligand charge transfer complexes this unique property may be adversely affected. We undertook UV-visible measurements in order to rule out this hypothesis. As seen in Figure 8, both bare and coated nanoceria have the same absorbance variation through the entire UV-visible region. This result is valuable because the robustness of the PPEG-Ceria nanoparticle sol that has been investigated in this work has direct impact for applications where anti-UV protection is needed.

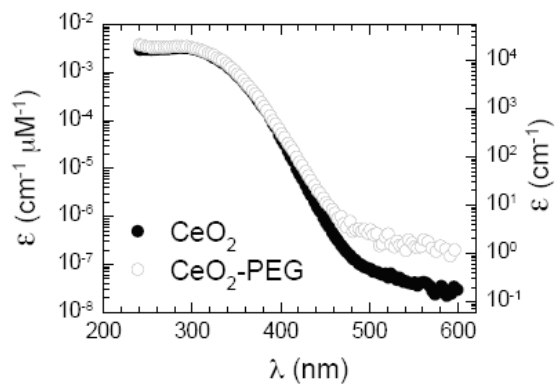

*Figure 8.* UV-visible spectra of bare and coated nanoceria.

Another test for the robustness of the surface modification and the ability to prepare stable concentrated solutions was evaluated by drying the hybrid nanosol. A second important matter is the possibility to handle powders rather than solutions to improve the shelf life expectancy or simply to prepare solutions in a different solvent.

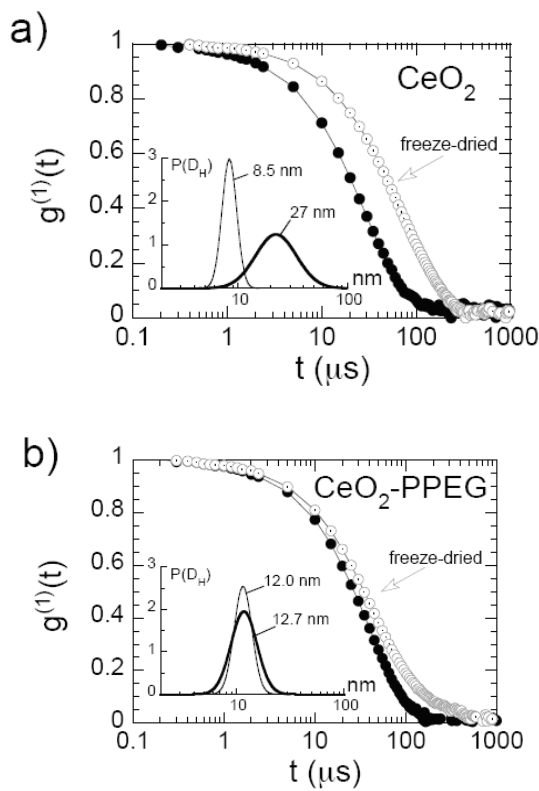

*Figure 9.* Re-dispersion in aqueous solutions of $CeO_2$ and PPEG-$CeO_2$ powders obtained by freeze-drying both stock solutions. a) Correlation functions of the original sol before and after freeze-drying and re-dispersion. The inset shows the derived hydrodynamic radii $D_H$ distribution. a) id. hybrid sol. For sake of comparison, the diffusion coefficients have been measured at a concentration of 0.1%. They have not been extrapolated to zero concentration leading to slightly smaller $D_H$ than in Table 1.

To further investigate this idea, we freeze-dried (Freezone model, Labconco, Kansas City, MO) bare and hybrid nanoceria solutions for 48 hours to remove any trace of the water. The phosphonated-PEG ceria solution was dialyzed (5K membrane) against DI water prior to freeze-drying, to remove any free PPEG oligomers present in the bulk.

Aliquots of the freeze dried powders so prepared when added to water showed a remarkable result that the hybrid nanosols appeared to readily redisperse to give clear solutions when compared to bare nanoparticles which evidenced a slightly milky appearance. Both powders were re-dispersed in aqueous solutions at pH = 1.5 for bare particles and DI water for stabilized particles and stirred overnight. Figure 9 shows the correlation functions measured with dynamic light scattering (at 90º) of the original and modified sol before and after re-dispersion. The distributions P(D) of the hydration diameters $D_H$ are also shown (inset Figure 9). After re-dispersion, the correlation function of the bare $CeO_2$ solution is clearly shifted toward longer times. As a result, the mean $D_H$ is shifted toward larger values together with a clear broadening of the distribution ($D_H$ = 26.8 ± 11.4 nm) compared to the small and slightly polydisperse particles in the original solution ($D_H$ = 8.5± 0.8 nm). Freeze drying the *reactive* bare $CeO_2$ particles would likely have led to some condensation between Ce-OH groups present on their surfaces giving rise to larger aggregates. In the case of *passivated* PPEG-$CeO_2$ particles, the freeze-drying process did not significantly change the original distribution indicating no change in the surface complexation of PPEG during the drying process. Both correlation functions are almost superimposed giving a very similar distribution of $D_H$ ($D_H$ = 12 ± 0.1.4 and $D_H$ = 12.7 ± 1.5). This redispersibility is only possible for compositions where $X < X_P$. This hybrid metal oxide nanopowder has ability to overcome strong van der Waals attractions by hydration/solvation of the PPEG corona yielding singlet nanocolloids while the formation of the barrier coating at the particle interface protects against interparticle chemical condensation. This has clear implications for the utility of nanoceria providing clear cost and processing advantages in allowing shipping of nanopowders rather than dilute solutions and in the preparation of stable concentrated solutions for bulk applications.

The sol stability and redispersability were further extended through the choice of the complexing oligomer backbone. The hybrid metal oxide powder is *also re-dispersible in certain organic solvents* like ethanol, acetone or chloroform (not shown here) which is not the case for bare particles. These results may easily be extended to other nanoparticle systems. The complexation of nanoceria with end-functional PPEG to create true redispersible nanopowders in aqueous or certain organic solvents provides the framework for designing a truly versatile hybrid metal oxide sol with clear utility in a range of applications.

## Concluding Remarks

The generation of robust and versatile phosphonated-PEG cerium oxide nanoparticles is highlighted through a detailed series of complementary experiments. From quantitative light and neutron scattering data we are able to obtain the main features of the $CeO_2$-PPEG hybrid core-shell nanostructure and to model the adsorption mechanism (NST model). The microstructure is described in terms of molecular weight $M_w$, hydrodynamic diameter $D_H$ and number of adsorbed oligomers per particle $N_{ads}$. The characteristic evolution of the Raleigh ratio as a function of the mixing ratio is successfully described in term of non specific adsorption of oligomers onto nanoparticles. The NST model evidenced the presence of a critical ratio $X_P$ at which the particles are fully covered with an average number of chains $N_{ads}$ = 270. Beyond that threshold, the organic coating progressively diminishes, increasing the sensitiveness of the sol toward destabilization. In addition, the measurement of the free energy of adsorption $\Delta G^{ads}$ (~ - 16 kT) has shown that electrostatic interaction is likely the main driving force the complexation which eventually increased the affinity of the phosphonate anchoring groups toward the $CeO_2$ surface.

The main specificities of the original sol are not altered during the functionalization phase leading to nanometer size cerium oxide particles covered with a well anchored layer of PEG chains. This solvating brush-like layer is sufficient to solubilize the particles and greatly expands the stability range of the original sol up to pH = 9. These results highlight the specificity of the terminus phosphonate group to create a robust metal oxide hybrid sol. Furthermore, tailoring the oligomer architecture with both a non interacting PEG tail and a mono-functional phosphonate head prevented colloidal bridging emphasizing the advantage of the current functionalization route over other ones using multi-sticker binding. Some direct benefits of such hybrid sols are finally put forward. After functionalization, they maintain a strong UV absorption capability, a very valuable characteristic for applications where anti-UV protection is needed. Moreover, after freeze-drying the hybrid particle solution, the powder was able to *re-disperse* in different solvents without losing any of its features conferring a great versatility and ease of use of the $CeO_2$-PPEG nanoparticles in dry or wet conditions. Beside their bulk properties described in this paper, the hybrid nanoparticles present some very interesting interfacial properties that will be discussed in a forthcoming publication.

## Materials and Methods

### Oligomers and nanoparticles

The phosphonated poly(oxyalkene) investigated in this work is a proprietary oligomer (poly(oxy-1 2 ethanediyl) alpha-(3-phosphonopropyl) omega hydroxyl or 3-phosphonopropyl alcohol ethoxylate-10 EO), produced by *Rhodia Inc.* and named hereafter phosphonated-PEG or PPEG (Figure 10). Titration curves with NaOH (1M) show the presence of two distinct p$K$as for this weak diacid at pKa$_1$ = 2.7 and pKa$_2$ = 7.8.

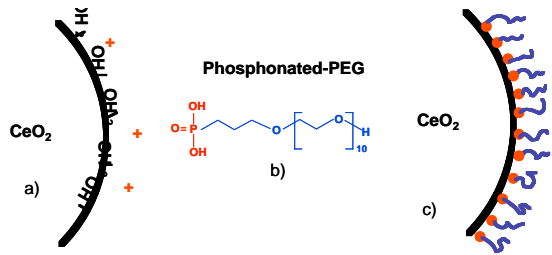

*Figure 10* : (a) Simplified sketch of the surface chemistry of cerium oxide nanoparticles: cationic protonated hydroxyl groups versus neutral hydroxyl groups. (b) Tailored phosphonated-PPEG chemical architecture for the adsorption onto nanoparticles: reactive phosphonate head + neutral PEG tail (c) Phosphonated-PEG corona around the particle: steric stabilization + PEG functionality.

The inorganic mineral oxide nanoparticle system investigated is a dispersion of cerium oxide nanocrystals, or nanoceria sol at pH 1.5. The synthetic procedure involves thermohydrolysis of an acidic solution of cerium-IV nitrate salt (Ce$^{4+}$ (NO$_3^-$)$_4$) at high temperature (70° C), that results in the homogeneous precipitation of a cerium oxide nanoparticle pulp (CeO$_2$ (HNO$_3$)$_{0.5}$ (H$_2$O)$_4$ )[37]. The size of the particles was controlled by addition of hydroxide ions during the thermohydrolysis. High resolution transmission electron microscopy has shown that the nanoceria (bulk mass density ρ = 7.1 g cm$^{-3}$) consists of isotropic agglomerates of 2 - 5 crystallites with typical size 2 nm and faceted morphologies. Wide-angle x-ray scattering has confirmed the crystalline fluorite structure of the nanocrystallites (see supplementary materials) [38]. Image analysis performed on cryo-TEM images of single nanoparticles have shown a polydispersity index s = 0.15 ± 0.03 for the particles (s is defined as the ratio between the standard deviation and the average diameter)[39].

As synthesized, the cerium oxide nanosols are stabilized by combination of long range electrostatic forces and short range hydration interactions (including strongly bound or condensed nitrate ions). At such pH, the ionic strength arises from the residual nitrate counter-ions present in the solution and acidic protons. This ionic strength around 0.045 M gives a Debye screening length $K_D^{-1}$ ~ 1.5 nm. An increase of the pH *or* ionic strength (> 0.45 M) results in a reversible aggregation of the particles, and destabilization of the sols leading eventually to a macroscopic phase separation. Though the change in pH apparently reduces the effective surface charge of the particles, it also reduces the range of the electrostatic repulsion. For this system, the destabilization of the sols occurs well below the point of zero charge of the ceria particles, pzc = 7.9 [28]. The bare nanoceria particles have a zeta potential ζ = + 30 mV and an estimated structural charge of $Q_{CeO_2}$ = + 300 e. The charges are compensated by nitrate anions in the Stern and diffuse layers surrounding each particle.

### Hybrid nanoparticles formulation

The pH of the PPEG solution is adjusted with reagent-grade nitric acid (HNO$_3$). Mixed solutions of nanoparticles (CeO$_2$) and phosphonate-PEG (PPEG) are prepared by simple mixing of dilute solutions prepared at the same concentration c (c = 0.1 - 1 wt. %) and same pH. This ensures that no aggregation of nanoparticules occurs due to pH or salinity gap. At pH = 1.4, 2.5 wt% of the phosphonate groups are ionized. The relative amount of each component is monitored by the volume ratio X, yielding for the final concentrations:

$$c_{CeO_2} = \frac{cX}{1+X} \ , \ c_{PPEG} = \frac{c}{1+X} \quad (7)$$

Ammonium hydroxide (NH$_4$OH) is used to adjust the pH of CeO$_2$-PPEG dispersions in the range of 1.5 to 10.

### Static and dynamic light scattering

Static and dynamic light scattering (SLS) measurements are performed on a BI-9000AT Brookhaven spectrometer (incident wavelength 488 nm) and on a Zetasizer Nano ZS from Malvern. Rayleigh ratios R and hydrodynamic diameters are measured as a function of the concentration c. R is obtained from the scattered intensity I(c):

$$R(q,c) = R_{std} \frac{I(c) - I_S}{I_{Tol}} \left( \frac{n}{n_{Tol}} \right)^2 \quad (8)$$

where $R_{std}$ and $n_{Tol}$ are the standard Rayleigh ratio and refractive index of toluene, $I_S$ and $I_{Tol}$ the intensities measured for the solvent and for the toluene in the same scattering configuration. Light scattering is used to determine the apparent molecular weight $M_{w,app}$ (the gyration radius $R_G$ was determined by neutron scattering) of colloids investigated here. In the regime of weak colloidal interactions, the Rayleigh ratio $R(c)$ is found to follow concentration dependence:

$$\frac{Kc}{R(q,c)} = \frac{1}{M_{w,app}} \left( 1 + \frac{q^2 R_g^2}{3} \right) + 2 A_2 c \quad (9)$$

where $K = 4\pi^2 n^2 (dn/dc)^2 / N_A \lambda^4$ is the scattering contrast coefficient ($N_A$ is the Avogadro number) and $A_2$ is the second virial coefficient. The refractive index increments dn/dc of the different solutions are measured using an BELLINGHAM & STANLEY model RFM840 refractive index detector in the range c = $10^{-3}$ to 1 wt. %. The values of the refractive index increments and K coefficients for the oligomers and nanoparticles solutions are shown in Table I. For the oligomers and the nanoparticles in the dilute concentration range (c < 1 wt. %), $qR_G \ll 1$, equation (9) reduces to

$$R(c) = K\,M_{w,app}\,c \qquad (10)$$

This expression emphasizes the fact that for small sizes (< $\lambda$/20), the Rayleigh ratio does not depend on the wave-vector in the window $6\times 10^{-4} - 4\times 10^{-3}$ Å$^{-1}$, characteristic for light scattering.

To accurately determine the size of the colloidal species, dynamic light scattering (DLS) was performed with concentration ranging from c = 0.01 wt. % – 1 wt. %. In this range, the diffusion coefficient varies according to: $D(c) = D_0(1 + D_2 c)$, where $D_0$ is the self-diffusion coefficient and $D_2$ is a virial coefficient of the series expansion. The sign of the virial coefficient, the type of interactions between the aggregates, either repulsive or attractive can be deduced. Here $D_2$ is positive (+ $10^{-7}$ cm$^2 \cdot$s$^{-2}$), indicating a repulsive interparticle interaction. From the value of D(c) extrapolated at c = 0 (noted $D_0$), the hydrodynamic radius of the colloids is calculated according to the Stokes-Einstein relation, $D_H = k_B T / 3\pi\eta_S D_0$, where $k_B$ is the Boltzmann constant, T the temperature (T = 298 K) and $\eta_S$ ($\eta_S$ = 0.89$\times 10^{-3}$ Pa.s) the solvent viscosity. The autocorrelation functions of the scattered light are interpreted using both the method of cumulants and the CONTIN fitting procedure provided by the instrument software.

### UV-visible absorption spectrometry

A UV-visible spectrometer (SmartSpecPlus from BioRad) is used to measure the absorbance of bare and coated nanoceria dispersion in water. The absorbance is related to the nanoparticle concentration by the Beer-Lambert expression:

$$\mathbf{Abs = \varepsilon\,l\,c_{NP} = -\log T \qquad (11)}$$

where $l$ ( = 1 cm), is the optical length of the cell, $c_{NP}$ the nanoparticle concentration, $\varepsilon$ the molar absorption coefficient (cm$^{-1}$ $\mu$M$^{-1}$), and T the transmission. In the following, the nanoparticle concentration will be expressed in wt. % or in $\mu$M of cerium oxide.

### Optical reflectometry

The adsorption isotherms $\Gamma(c)$ of PPEG oligomers onto a cerium oxide model surface is monitored using optical reflectometry[40,41]. Fixed angle reflectometry measures the reflectance at the Brewster angle on the flat substrate. A linearly polarized light beam is reflected by the surface and subsequently split into a parallel and a perpendicular component using a polarizing beam splitter. As material adsorbed at the substrate-solution interface, the intensity ratio S between the parallel and perpendicular components of the reflected light is varied. The change in S is related to the adsorbed amount through:

$$\Gamma(t) = \frac{1}{A_s}\,\frac{S(t) - S_0}{S_0} \qquad (12)$$

where $S_0$ is the signal from the bare surface prior to adsorption. $\Gamma(c)$ is constructed by taking the plateau value $\Gamma_{\text{Plateau}}$ of a given $\Gamma(t)$ curve at different concentration c (c = 0.001 to 0.1 wt. %). A complete description of the stagnation point adsorption reflectometry device can be found in several references[41-46].

*CeO$_2$ model surfaces.* A thin layer of polystyrene $\sim$ 100 nm is deposited on top of an HMDS (hexamethyldisilizane) functionalized silicon wafer by spin-coating a toluene solution (25 g/l) at 5000 rpm[47]. The surface is then dipped in a nanoceria solution (0.1 wt. %) containing 0.1 M NaNO$_3$ overnight. This results in the formation of a well packed nanoceria monolayer on top of the PS surface as seen by AFM imaging (not shown here). Bare particle adsorption on plastic surfaces will be described in a forthcoming publication. It should be noted that the receding water contact angle $\theta_r$ on such model nanoceria surface was below 15° in contrast with a $\theta_r$ = 85 ° for the original PS surface.

### Small-Angle Neutron Scattering

Small-Angle Neutron Scattering (SANS) spectra are measured on PAXY spectrometer (Laboratoire Leon Brillouin-LLB, Saclay, France). Two configurations are used (*D* = 1.35 and 6.70 m, both at $\lambda$ = 6 Å), covering a *q*-range from $5\times 10^{-3}$ Å$^{-1}$ to 0.2 Å$^{-1}$. Exposure times of 2 h and 1 h for the small and large angle configuration respectively are necessary to obtain a good statistics. Raw data are radially averaged. Standard corrections for sample volume, neutron beam transmission, empty cell signal subtraction, and detector efficiency have been applied to get the scattered intensities in absolute scale using standard LLB software.

**Supporting Information Available:**
WAXS and HR-TEM data for bare $CeO_2$ particles. This material is available free of charge via the Internet at http://pubs.acs.org.